\documentclass[a4paper,UKenglish,cleveref, autoref, thm-restate]{lipics-v2021}

\usepackage{todonotes}
\usepackage{xspace}
\usepackage{stmaryrd} 
\usepackage{algorithm}
\usepackage{algpseudocodex}
\usepackage{caption}
\usepackage{subcaption}
\usepackage{booktabs}
\nolinenumbers

\usepackage{array}
\usepackage{hhline}
\newcolumntype{L}[1]{>{\raggedright\let\newline\\\arraybackslash\hspace{0pt}}m{#1}}
\newcolumntype{C}[1]{>{\centering\let\newline\\\arraybackslash\hspace{0pt}}m{#1}}
\newcolumntype{R}[1]{>{\raggedleft\let\newline\\\arraybackslash\hspace{0pt}}m{#1}}

\title{SMT with Uninterpreted Functions and Monotonicity Constraints in Systems Biology} 

\author{Ondřej {Huvar}}{Masaryk University, Czechia}{xhuvar@fi.muni.cz}{}{}
\author{Martin {Jonáš}}{Masaryk University, Czechia}{martin.jonas@mail.muni.cz}{https://orcid.org/0000-0003-4703-0795}{}
\author{Samuel {Pastva}}{Masaryk University, Czechia}{daemontus@mail.muni.cz}{https://orcid.org/0000-0003-1993-0331}{}

\authorrunning{O. Huvar and M. Jonáš and S. Pastva}

\Copyright{Ondřej Huvar and Martin Jonáš and Samuel Pastva}



\category{} 

\relatedversion{} 

\EventEditors{John Q. Open and Joan R. Access}
\EventNoEds{2}
\EventLongTitle{42nd Conference on Very Important Topics (CVIT 2016)}
\EventShortTitle{CVIT 2016}
\EventAcronym{CVIT}
\EventYear{2016}
\EventDate{December 24--27, 2016}
\EventLocation{Little Whinging, United Kingdom}
\EventLogo{}
\SeriesVolume{42}
\ArticleNo{23}

\newcommand{\N}{\mathbb{N}}
\newcommand{\vect}[1]{\overline{#1}}
\newcommand{\subst}[3]{\ensuremath{{#1}[#2 \leftarrow #3]}}
\newcommand{\arity}[1]{\ensuremath{\mathrm{arity}(#1)}}

\newcommand{\vectIndex}[2]{\ensuremath{{#1}_{#2}}}
\newcommand{\UF}[0]{\ensuremath{\mathrm{UF}}}
\newcommand{\LIA}[0]{\ensuremath{\mathrm{LIA}}}

\newcommand{\terms}[1]{\ensuremath{\mathit{terms}(#1)}}

\newcommand{\solveEager}{\textsc{SolveEager}\xspace}
\newcommand{\smt}{\textsc{smt}\xspace}
\newcommand{\dnf}{\textsc{dnf}\xspace}
\newcommand{\asp}{\textsc{asp}\xspace}
\newcommand{\bdd}{\textsc{bdd}\xspace}
\newcommand{\bma}{\textsc{bma}\xspace}
\newcommand{\bbm}{\textsc{bbm}\xspace}

\begin{document}

\maketitle

\begin{abstract}

  The theory of uninterpreted functions is a key modeling tool for
  systems with unknown or abstracted components.  Some domains such as
  systems biology impose further restrictions regarding monotonicity
  on these components, requiring specific inputs to have a
  consistently positive or negative effect on the output.  In this
  paper, we tackle the \emph{model inference} problem for biological
  systems by applying the theory of uninterpreted functions with
  monotonicity constraints. We compare the performance of naive
  quantified encodings of the problem and the performance of the
  existing approach based on eager quantifier instantiation, which is
  based on the fact that a finite set of quantifier-free monotonicity
  lemmas is sufficient to encode the monotonicity of uninterpreted
  functions. Additionally, we consider a lazy variant of the approach
  that introduces the monotonicity lemmas on demand.

  We evaluate the \smt-based approach to model inference using a large
  collection of systems biology benchmarks. The results demonstrate
  that the instantiation-based encodings significantly outperform
  quantified encodings, which typically struggle with large function
  arities and complex instances. As the key result, we show that our
  approach based on \smt with uninterpreted functions and monotonicity
  constraints significantly outperforms state-of-the-art
  domain-specific tools used in systems biology, such as the
  \asp-based Bonesis and the \bdd-based AEON.
\end{abstract}

\section{Introduction}

The theory of uninterpreted functions is a key modeling tool for systems with
unknown or abstracted components. Some domains, including systems biology,
impose further \emph{monotonicity} constraints on these components,
requiring specific inputs to have a consistently positive or negative effect on
the output. The monotonicity (or anti-monotonicity) of function arguments is a
common constraint in models of physical---particularly
biological---systems~\cite{kadelka2024meta} due to the mechanistic nature of
their underlying processes. For instance, a single chemical reaction either \emph{produces}
or \emph{consumes} a specific compound overall. Similarly, a catalyst or
inhibitor either \emph{accelerates} or \emph{slows down} such a reaction. While
there are rare cases of true non-monotonic influence in nature, monotonicity remains
a standard assumption in dynamical models of physical
systems~\cite{chevalier2025data,gerber2025reasoning,kadelka2024meta}.

There are existing techniques for solving satisfiability over theories
extended with uninterpreted functions subject to monotonicity
constraints. A straightforward approach for dealing with these
constraints relies on direct encoding using universally quantified
formulas. More notably, Sofronie-Stokkermans et al.~have proven
that the extension of a theory with uninterpreted functions and monotonicity
constraints is \emph{local}~\cite{sofronie2007automated,sofronie2014ieee}. As a
consequence, a fixed finite set of quantifier instances of
monotonicity axioms is sufficient to encode the monotonicity. These
instances can be derived directly from occurrences of uninterpreted function
symbols in the input formula~\cite{sofronie2007automated}.

This paper focuses on the use of uninterpreted functions with
monotonicity constraints in logic-based discrete models, such as
Boolean~\cite{kauffman1969homeostasis} and
Thomas~\cite{thomas1991regulatory} networks. These frameworks are
prominent in systems biology as predictive models of emergent
phenomena~\cite{rozum2024boolean} and are widely applied to model gene
regulation, cell-to-cell signaling, and metabolic
pathways~\cite{kadelka2024meta,pastva2023repository}. They consist of
discrete variables (binary in Boolean networks, multi-valued in Thomas
networks) that represent the discretized concentrations of biochemical
species. Each variable is governed by an \emph{update function}, which
maps the current system state to the variable's level in the next time
step.

\begin{figure}
	\centering

	\begin{minipage}{0.45\linewidth}
		\centering
		\includegraphics[width=0.6\linewidth]{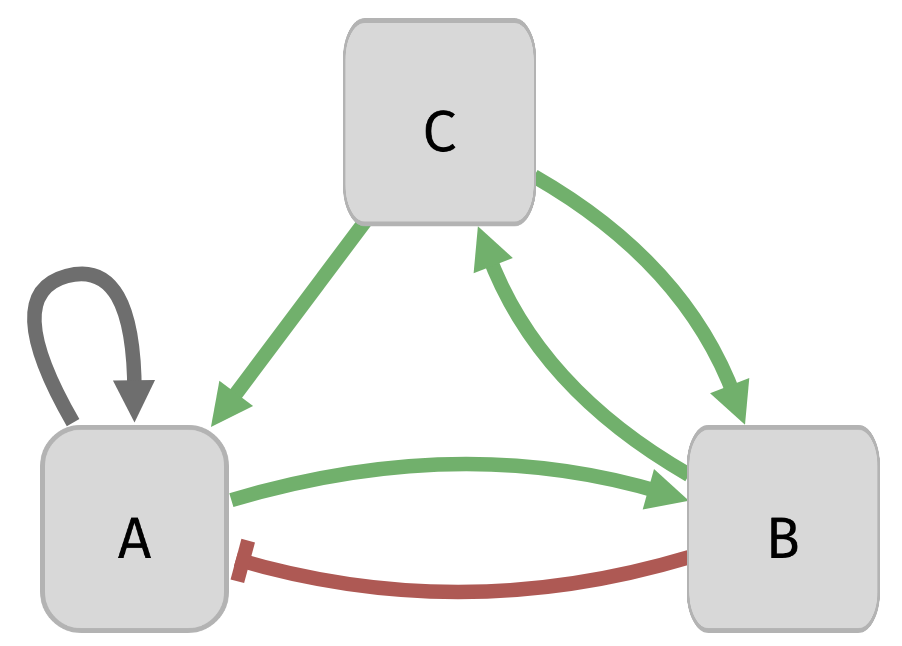}
		\vspace{8pt}

		\footnotesize
		$f_a(a, b, c)$\\$f_b(a, c)$\\$f_c(b)$
	\end{minipage}
	\hspace{6pt}
	\begin{minipage}{0.45\linewidth}
		\footnotesize
		\begin{itemize}
			\item Argument monotonicity of $f_a$:
			\begin{itemize}
				\item unrestricted in $a$;
				\item anti-monotone in $b$;
				\item monotone in $c$;
			\end{itemize}
			\item Argument monotonicity of $f_b$:
			\begin{itemize}
				\item monotone in $a$ and $c$;
			\end{itemize}
			\item Argument monotonicity of $f_c$:
			\begin{itemize}
				\item monotone in $b$;
			\end{itemize}
			\item State $(0,0,0)$ is a fixed-point;
			\item State $(0,1,1)$ is a fixed-point;
			\item State $(1,2,2)$ is a fixed-point.
		\end{itemize}
	\end{minipage}
	\vspace{8pt}

	\begin{minipage}[t]{0.45\linewidth}
		\footnotesize
		(a) Influence graph of a hypothetical biological system with unknown update functions represented by $f_a$, $f_b$, and $f_c$.
		Green arrows depict positive, red arrows negative, and gray arrows unrestricted regulations.
	\end{minipage}
	\hspace{6pt}
	\begin{minipage}[t]{0.45\linewidth}
		\footnotesize
		(b) Restrictions imposed on the biological system either through empirical observations (fixed points)
		or via mechanistic assumptions (argument monotonicity).
		All function arguments are required to be essential (i.e., influence the function's output).
	\end{minipage}
	\hspace{6pt}

	\caption{A simple example depicting an inference
		problem of a three-component system. It consists of three genes
		($a$, $b$, and $c$) with integer domains and six expected
		regulations (dependencies) between said genes, some with
		monotonicity constraints. Finally, its long-term behavior is
		defined by three observations of fixed points.}
	\label{fig:inference-example}
\end{figure}

A central challenge in logic-based modeling is \emph{model inference}:
the task of identifying a concrete model of the system that satisfies
all empirical observations and mechanistic assumptions. While these
constraints can vary, observations most frequently represent observed
steady states (i.e., fixed points) of the system. Mechanistic
assumptions, meanwhile, typically define the essentiality and monotonicity
of regulations between variables, which together constitute the network's influence
graph.

Figure~\ref{fig:inference-example} depicts a simple inference
problem. The system has three integer-valued variables: $a$, $b$, and
$c$. The precise update functions that compute the next state of the
system are unknown, but it has been observed that $a$ is influenced
by all three variables, with the effect of $b$ being negative
and the effect of $c$ positive. Furthermore, $b$ is positively influenced
by both $a$ and $c$, while $c$ is positively influenced by
$b$. It has also been experimentally observed that the configurations
$(0,0,0)$, $(0,1,1)$, and $(1,2,2)$ are fixed points of the
system. For example, when the system is in the state $(1,2,2)$, it must
remain in $(1,2,2)$ in the next state. One admissible solution for
this inference problem is given by:
\begin{align*}
  f_a(a, b, c) &= \begin{cases}
    \max(0, 3-b), &\text{if } a=1, \\
    \max(0, c-b), & \text{otherwise},
  \end{cases} \\
  f_b(a, c) &= \max(a, c), \\
  f_c(b) &= b.
\end{align*}
This set of update functions clearly satisfies that $f_a(a, b, c)$
is anti-monotone in $b$ and monotone in $c$, and that $f_b(a, c)$ and
$f_c(b)$ are monotone in all their arguments. Moreover, the update
functions satisfy the observed fixed points:
\begin{enumerate}
\item For the state $(0,0,0)$, the next state is
  $(f_a(0,0,0), f_b(0,0), f_c(0))$, which is $(0,0,0)$.
\item For the state $(0,1,1)$, the next state is
  $(f_a(0,1,1), f_b(0,1), f_c(1))$, which is $(0,1,1)$.
\item For the state $(1,2,2)$, the next state is
  $(f_a(1,2,2), f_b(1,2), f_c(2))$, which is $(1,2,2)$.
\end{enumerate}

\medskip

\noindent \emph{Contribution.} We show that the model inference
problem maps naturally to \smt, with unknown update functions directly
encoded as uninterpreted functions subject to additional constraints
derived from fixed-point observations and the regulations of the
influence graph. The mechanistic assumptions regarding regulatory monotonicity
translate naturally into monotonicity constraints on these
uninterpreted functions. Using the resulting \smt problems, we compare
the performance of naive quantified encodings against the previously proposed approach~\cite{sofronie2007automated} based on eager quantifier instantiation
Additionally, we introduce a lazy
variant of the instantiation approach that introduces monotonicity lemmas on demand.

We evaluate these \smt-based approaches to model inference using a
comprehensive suite of benchmarks from systems biology, comprising
8,381 and 465 instances derived from existing Boolean and integer-based
models, respectively, as well as 144 large instances from biological
datasets without a known ground-truth model. Our results demonstrate
that the instantiation-based procedures significantly outperform the
naive quantified encoding, particularly on complex or high-arity
problems. As a key result, we show that our \smt-based approach
significantly outperforms state-of-the-art tools used in systems
biology, such as the \asp-based Bonesis and the \bdd-based AEON.
Moreover, unlike these tools, our approach enables the analysis of non-Boolean models.

\medskip

\noindent \emph{Paper structure.} The paper is structured as
follows. Sections~\ref{sec:preliminaries} and \ref{sec:monotonicity}
briefly recall the logical preliminaries and monotonicity
constraints in the context of \smt. In
Section~\ref{sec:inference-encoding}, we describe the encoding of the
inference problem using uninterpreted functions with monotonicity
constraints. Section~\ref{sec:solving} covers existing approaches
to solving such constraints and introduces the lazy instantiation-based algorithm.
Finally, Section~\ref{sec:experiments} presents the experimental evaluation,
and Section~\ref{sec:conclusions} concludes the paper.

\medskip

\noindent \emph{Related work.} Several existing approaches tackle the
model inference problem in the presence of monotonicity constraints without
relying on \smt with uninterpreted functions. While some of these address model
inference through statistical methods or genetic
programming~\cite{puvsnik2022review}, a significant effort is dedicated toward
model inference via automated reasoning. Across these methods, efficiently
enforcing monotonicity remains a persistent challenge. The tool
RE:IN~\cite{yordanov2016method,yordanov2023reasoning} encodes inference problems
into \smt \emph{without} uninterpreted functions; to guarantee monotonicity, it
restricts the search space to a limited, syntactically defined subset of update
functions. Bonesis~\cite{chevalier2024bonesis,chevalier2025data} and
Caspo~\cite{guziolowski2013exhaustively} support only Boolean models and rely on
Answer Set Programming (\asp) with exhaustive Disjunctive Normal Form (\dnf)
encodings. Since the size of a \dnf encoding grows exponentially with function
arity, Bonesis offers an optimization that limits the number of permitted \dnf
cubes, thereby trading completeness for scalability. Finally,
AEON~\cite{benevs2022aeon,benevs2023boolean} is also limited to Boolean models
and employs Binary Decision Diagrams ({\bdd}s) to exhaustively represent the
function space. The \bdd encoding---for each update function and for each
combination of its input values---introduces one \bdd variable representing the
result of the function and imposes the monotonicity constraints directly at the
\bdd level. This induces an exponential blowup, restricting the tool's
applicability to functions of low arity. Notably, neither of these tools support
integer-valued models.

\section{Preliminaries}
\label{sec:preliminaries}

Let $A$ be a set. For a vector $\vect{x} \in A^n$, we denote by
$\vectIndex{x}{i}$ the value at its $i$-th position. For any vector
$\vect{x} \in A^n$, element $b \in A$, and index $1 \leq i \leq n$, we denote by
$\subst{\vect{x}}{i}{b}$ the vector that agrees with $\vect{x}$ at all positions
except at position $i$, where it has the value $b$.

Let $(A, \leq)$ be a partially ordered set. The ordering is called \emph{total}
if for all $a, b \in A$, we have either $a \leq b$ or $b \leq a$. For each
$n \in \N$ and $1 \leq i \leq n$, a function $f \colon A^n \rightarrow A$ is
called \emph{monotone in the $i$-th argument} if for all $\vect{a} \in A^n$ and
$b \in A$, we have that $\vectIndex{a}{i} \leq b$ implies
$f(\vect{a}) \leq f(\subst{\vect{a}}{i}{b})$. The function is
\emph{anti-monotone in the $i$-th argument} if the same assumptions imply
$f(\vect{a}) \geq f(\subst{\vect{a}}{i}{b})$.

We work in the context of standard first-order logic with
equality. We briefly recall the notation and concepts that are
important for this paper and refer the reader to standard textbooks on
the topic for further details~\cite{Enderton72,BT18}. As is standard in \smt, we
consider only quantifier-free formulas, unless explicitly stated
otherwise.

A signature $\Sigma$ is a set of function symbols and predicate symbols, denoted
$\Sigma^F$ and $\Sigma^P$, respectively. Each $f \in \Sigma^F$ and
$p \in \Sigma^P$ has an associated \emph{arity}, denoted $\arity{f} \in \N$ and
$\arity{p} \in \N$, respectively. The sets of \emph{($\Sigma$-)terms} and
\emph{($\Sigma$-)formulas} are defined in the standard way. We denote by
$\terms{\varphi}$ the set of all subterms of a formula $\varphi$.

A \emph{($\Sigma$-)structure} $\mathcal{A}$ is a pair consisting of a
non-empty set $A$ (the domain) and a mapping $(\_)^\mathcal{A}$ that to each
function symbol $f$ of arity $n$ assigns a function
$f^\mathcal{A} \colon A^n \rightarrow A$ and to each predicate symbol
$P$ of arity $n$ assigns a relation $P^\mathcal{A} \subseteq
A^n$.
The
satisfaction relation $\mathcal{A} \models \varphi$ between a
structure $\mathcal{A}$ and a formula $\varphi$ is defined in the
standard way. A \emph{($\Sigma$-)theory} $T$ is a non-empty class of
$\Sigma$-structures. A formula $\varphi$ is \emph{$T$-satisfiable} (or
\emph{satisfiable modulo $T$}) if it is satisfied by some structure
$\mathcal{A} \in T$. It is \emph{$T$-unsatisfiable} otherwise. A formula is
\emph{$T$-valid} if it is satisfied by all structures in $T$. A function
symbol $f \in \Sigma^F$ is called \emph{uninterpreted} in the theory
$T$ if the theory does not impose any restrictions on its
interpretation. That is, if two $\Sigma$-structures
$\mathcal{A}$ and $\mathcal{A}'$ differ only in the interpretation of $f$,
we have $\mathcal{A} \in T \Leftrightarrow \mathcal{A}' \in T$.

The examples in this paper are over the theory $T$ of linear integer
arithmetic (\LIA), whose structures have the domain $\mathbb{Z}$ and the
standard interpretation of arithmetic function and predicate
symbols. However, the approach presented here is not limited to any
particular theory $T$. We only assume that we are working over a
theory $T$ for which satisfiability of quantifier-free formulas is
decidable.

\section{Monotonicity Constraints}
\label{sec:monotonicity}

In this section, we formalize the monotonicity constraints. We first
define \emph{a monotonicity specification}, which states, for each uninterpreted
function symbol, which of its formal parameters are monotone,
which are anti-monotone, and which are unrestricted. We then define
when a given formula $\varphi$ is satisfiable in a given logical
theory with respect to a given monotonicity specification.
Our definition is compatible with other notions of monotonicity constraints
appearing in the literature~\cite{sofronie2007automated}.

In the rest of the section, let $T$ be a fixed theory over a signature $\Sigma$
that contains a binary predicate symbol $\leq$ corresponding to a \emph{total}
ordering relation\footnote{The encodings of monotonicity used later in the paper
  are correct even with weaker assumptions on the ordering. The quantified
  encoding works for arbitrary partial orderings and the encoding based on eager
  quantifier instantiation works for arbitrary lattices and bounded
  semilattices~\cite{sofronie2007automated,sofronie2014ieee}.}. Formally, there
is a predicate symbol ${\leq} \in \Sigma^P$ of arity $2$ such that for all
structures $\mathcal{A} = (A, (\_)^\mathcal{A}) \in T$, the interpretation
$\leq^\mathcal{A}$ of this symbol is a total ordering relation on $A$.

\begin{definition}[Monotonicity specification]
  \label{def:monotonicity}
  Let $f \in \Sigma^F$ be a function symbol of arity~$n$. A~\emph{monotonicity
    specification for $f$} is a pair $(f_m, f_a)$ where
  $f_m \subseteq \{1, \dots, n\}$ and $f_a \subseteq \{1, \dots, n\}$ are respectively the
  indices of \emph{monotone} and \emph{anti-monotone} arguments of
  $f$. A $\Sigma$-structure $\mathcal{A}$ satisfies the monotonicity
  specification $(f_m, f_a)$, written
  $\mathcal{A} \models (f_m, f_a)$, if:
  \begin{itemize}
  \item $f^\mathcal{A}$ is monotone in the $i$-th argument for each $i \in f_m$,
  \item $f^\mathcal{A}$ is anti-monotone in the $i$-th argument for each $i \in f_a$.
  \end{itemize}

  \noindent A mapping $M$ that assigns a monotonicity specification
  $M(f)$ to each uninterpreted function symbol $f \in \Sigma^F$ is
  called a \emph{monotonicity specification for $\Sigma$}. The
  specification $M$ is satisfied by a structure $\mathcal{A}$,
  written $\mathcal{A} \models M$, if for each uninterpreted function
  symbol $f$, we have $\mathcal{A} \models M(f)$.
\end{definition}

\begin{definition}
  A formula $\varphi$ is satisfiable modulo the theory $T$ with
  monotonicity specification $M$ (or $(T,M)$-satisfiable) if there is
  a $\Sigma$-structure $\mathcal{A} \in T$ such that
  $\mathcal{A} \models \varphi$ and $\mathcal{A} \models M$.
\end{definition}

As a notational shorthand, if the monotonicity specification is clear from the
context, we write simply $f_m$ for the monotone indices of $f$ and $f_a$ for
its anti-monotone indices.

Notably, this definition restricts the monotonicity specification only to
\emph{uninterpreted} function symbols. This is our intended
use-case, and the algorithms used later in the paper rely on this
assumption. The definition of a monotonicity specification could, in principle,
be extended to support interpreted functions as well, but enforcing it
would require different techniques.

\begin{example}
  \label{ex:running}
  Consider the theory $\LIA$ with two uninterpreted function symbols $f$ and $g$
  of arities 2 and 1, respectively. Further, consider the formula
  \begin{align*}
    \varphi ~\equiv~ ( f(c_1, 2) = 4 ) \wedge (f(c_1+5, 0) = c_2) \wedge (g(c_2) < g(4)).
  \end{align*}
  Let $M$ be the following monotonicity specification:
  \begin{align*}
    M(f) &= (\{ 1 \}, \{ 2 \} ), \\
    M(g) &= (\{ 1 \}, \emptyset ),
  \end{align*}
  i.e., the function $f$ is monotone in its first argument and anti-monotone in
  its second argument, and the function $g$ is monotone in its only argument.
  Then $\varphi$ is not satisfiable modulo $\LIA$ with the monotonicity
  specification $M$. The monotonicity of $f$ in the first argument
  implies $f(c_1, 2) \leq f(c_1+5, 2)$, and its anti-monotonicity in the second implies
  $f(c_1+5, 2) \leq f(c_1+5, 0)$. Together, these imply
  $4 = f(c_1, 2) \leq f(c_1+5, 0) = c_2$. Consequently, the monotonicity of $g$ implies $g(4)
  \leq g(c_2)$, which contradicts the last conjunct of $\varphi$.

  Now consider a similar monotonicity specification $M'$, where $f$ is not
  constrained to be anti-monotone, i.e.,
  $M'(f) = M'(g) = (\{ 1 \}, \emptyset )$. The formula $\varphi$ is now
  satisfiable modulo $\LIA$ with the monotonicity specification $M'$, because
  the model $\mathcal{A}$ with $c_1^\mathcal{A} = 4$, $c_2^\mathcal{A} = 0$,
  $g^\mathcal{A}(x) = x$, and $f^\mathcal{A}(x, y) = x$ for $y = 2$ and
  $f^\mathcal{A}(x, y) = 0$ otherwise satisfies both $\varphi$ and the
  monotonicity specification $M'$. Note that the monotonicity specification $M'$
  does not constrain the behavior of $f^\mathcal{A}$ in its second argument; the
  function can behave differently for each value of $y$, but it must be monotone
  in $x$ for each fixed $y$.
\end{example}

\section{SMT Encoding of Logic-based Model Inference}
\label{sec:inference-encoding}

We consider a standard variant of the logic-based model inference problem
over a set of $n$ indexed variables, $V = \{ v_1, \ldots, v_n \}$.
Each variable $v_i \in V$ has an associated domain $D_{v_i}$,
which is either Boolean ($D_{v_i} = \{ 0, 1 \}$) or a larger integer interval (e.g., $D_{v_i} = \{ 0, \ldots, 5 \}$).
Alternatively, the cases of $D_{v_i} = \mathbb{N}$ or $D_{v_i} = \mathbb{Z}$ are also possible, but are
very rare in practical biological applications, since concentrations of
biochemical species often have a finite number of distinct levels.
Models whose variables are restricted entirely to Boolean domains correspond to Boolean networks,
while those containing non-Boolean but bounded variables are known as multi-valued (or Thomas) networks.

The model structure is constrained by a directed influence graph $(V, E)$,
where edges $E \subseteq V \times V$ represent \emph{regulations}
(i.e., $v_i$ regulates $v_j$ if and only if $(v_i, v_j) \in E$).
This graph dictates the signature of the uninterpreted functions in our \smt query:
for each $v_i$, we define an uninterpreted function $f_{v_i}$ representing its update function,
with an arity and domain matching its regulators.

The inference problem imposes three types of constraints on these uninterpreted functions.\footnote{Here and in the rest of the section, we present the constraints as explicitly existentially quantified to avoid introducing fresh constant names for each constraint. However, the existentially quantified variables should be replaced by fresh constants symbols in the real application.}
\begin{itemize}
	\item \textbf{Monotonicity:} As part of the input, the method assumes a monotonicity specification~$M$. This is directly expressed as a monotonicity specification from Definition~\ref{def:monotonicity}.

	\item \textbf{Essentiality:} A regulation $(v_i, v_j) \in E$ may be declared \emph{essential}, meaning the regulator $v_i$ must have an observable impact on $v_j$. Assuming without loss of generality that $v_i$ is the first of $k$ regulators for $v_j$, we encode essentiality as a constraint $\eta_{v_j}^{v_i}$ defined as:
	\begin{equation*}
		\eta_{v_j}^{v_i} = \exists x, y, z_2, \ldots, z_k .\, f_{v_j}(x, z_2, \ldots, z_k) \neq f_{v_j}(y,z_2, \ldots, z_k).
	\end{equation*}

	\item \textbf{Fixed-point observations:} Finally, a fixed-point observation is a partial function assigning constant values to a subset of model variables, asserting the existence of a matching fixed-point in the model's state space. Let $\mathcal{F}$ be a set of such assignments $(v_i, d_i)$ where $v_i \in V$ and $d_i \in D_{v_i}$. The variables that are not assigned by $\mathcal{F}$ can be assigned arbitrarily. If $r_1^{(v_i)}, \ldots, r_k^{(v_i)}$ denote the indices of the $k$ regulators of $v_i$, the fixed-point constraint $\tau_{\mathcal{F}}$ is defined as follows:
	\begin{equation*}
		\tau_{\mathcal{F}} = \exists x_1, \ldots, x_n .\, \bigwedge_{1 \leq i \leq n} x_i = f_{v_i}(x_{r_1^{(v_i)}}, \ldots, x_{r_k^{(v_i)}}) \wedge \bigwedge_{(v_i, d_i) \in \mathcal{F}} x_i = d_i.
	\end{equation*}
\end{itemize}

Finally, for each variable or uninterpreted function with a bounded
domain, we introduce a constraint asserting these bounds. For Boolean
variables, the encoding uses propositional variables instead of
bounded integers. Values of propositional variables are naturally
ordered by $\mathit{false} < \mathit{true}$, which can be expressed
in propositional logic.

We can also apply two key formula simplifications. First, for
essential regulations where $v_i$ is Boolean, the quantified variables
$x$ and $y$ are directly instantiated with the constants $1$ and $0$.
Second, in fixed-point constraints, we propagate the observation
$x_i = d_i$, substituting $x_i$ with the constant $d_i$ in all
function applications.  While straightforward, we observe that these
optimizations can have a non-trivial impact on the overall
performance.

\begin{example}
	\label{ex:inference-encoding}

	Recall the inference problem that we described in Figure~\ref{fig:inference-example}.
	Formally, it prescribes three uninterpreted functions $f_a$, $f_b$, and $f_c$,
	with $\arity{f_a} = 3$, $\arity{f_b} = 2$, and $\arity{f_c} = 1$. Furthermore, the problem carries
	a monotonicity specification $M$ such that $M(f_a) = (\{3\}, \{2\})$, $M(f_b) = (\{1,2\}, \emptyset)$,
	and $M(f_c) = (\{1\}, \emptyset)$. Since each regulation here is expected to be essential, we consider
	six essentiality constraints:
	\begin{align*}
		\eta_{a}^{a} &~\equiv~ \exists x,y, z_2, z_3 .\, f_a(x,z_2,z_3) \neq f_a(y,z_2,z_3) & \eta_{b}^{a} &~\equiv~ \exists x,y, z .\, f_b(x,z) \neq f_b(y,z)\\
		\eta_{a}^{b} &~\equiv~ \exists x,y, z_1, z_3 .\, f_a(z_1, x,z_3) \neq f_a(z_1, y,z_3) & \eta_{b}^{c} &~\equiv~ \exists x,y, z .\, f_b(z,x) \neq f_b(z,y) \\
		\eta_{a}^{c} &~\equiv~ \exists x,y, z_1, z_2 .\, f_a(z_1,z_2, x) \neq f_a(z_1,z_2,y) & \eta_{c}^{b} &~\equiv~ \exists x,y .\, f_c(x) \neq f_c(y)
	\end{align*}
	To assert the expected fixed points $\mathcal{F}_1 = \{ (a, 0), (b, 0), (c, 0) \}$,
	$\mathcal{F}_2 =  \{ (a, 0), (b, 1), (c, 1) \}$, and $\mathcal{F}_3 =  \{ (a, 1), (b, 2), (c, 2) \}$,
	we also consider the following fixed-point constraints. Here, we applied the aforementioned
	propagation of observed values. Because we have complete information about
	each fixed point, this entirely eliminates the existentially quantified variables:
	\begin{align*}
		\tau_{\mathcal{F}_1} &~~\equiv~~ f_a(0, 0, 0) = 0 \wedge f_b(0, 0) = 0 \wedge f_c(0) = 0,\\
		\tau_{\mathcal{F}_2} &~~\equiv~~ f_a(0, 1, 1) = 0 \wedge f_b(0, 1) = 1 \wedge f_c(1) = 1,\\
		\tau_{\mathcal{F}_3} &~~\equiv~~ f_a(1, 2, 2) = 1 \wedge f_b(1, 2) = 2 \wedge f_c(2) = 2.
	\end{align*}
	Overall, the encoding of our inference problem is then the following formula $\varphi$ with the monotonicity specification $M$:
	\begin{equation*}
      \varphi ~~\equiv~~ (\eta_{a}^{a} \wedge \eta_{a}^{b} \wedge \eta_{a}^{c} \wedge \eta_{b}^{a} \wedge \eta_{b}^{c} \wedge \eta_{c}^{b}) \wedge (\tau_{\mathcal{F}_1} \wedge \tau_{\mathcal{F}_2} \wedge \tau_{\mathcal{F}_3}).
    \end{equation*}

	If we were to additionally require that the three system variables $a$, $b$, and $c$
	have bounded domains (e.g., $\{ 0, \ldots, 3 \}$), we must also introduce
	these bounds as separate constraints on every $f_a(\vect{x})$, $f_b(\vect{y})$,
	and $f_c(\vect{z})$ that appear in $\terms{\varphi}$, as well as on any
	existentially quantified atoms in the $\eta_{v_j}^{v_i}$ and $\tau_{\mathcal{F}_i}$ constraints.
\end{example}

\section{Solving SMT with Monotonicity Constraints}
\label{sec:solving}

In this section, we discuss in detail two existing techniques for
solving quantifier-free formulas with monotonicity constraints and
introduce a lazy approach based on generation of monotonicity lemmas
on demand.

\subsection{Quantified Encoding}
\label{sec:quantified-encoding}

A straightforward approach is to translate each formula $\varphi$ with
a monotonicity specification $M$ into an equivalent formula $\varphi^\mathit{quant}_M$
by directly encoding the monotonicity specification as a quantified formula
$\psi$ and defining $\varphi^\mathit{quant}_M = \varphi \wedge \psi$.

In particular, the formula $\psi$ can be defined as $\psi = \bigwedge_{f \in
  \Sigma^F} \bigwedge_{i \in (f_m \cup f_a)} \psi^f_i$, where
\[
  \psi^f_i =
  \begin{cases}
    \forall \vect{x} \forall y \left( \vectIndex{x}{i} \leq y \rightarrow f(\vect{x}) \leq f(\subst{\vect{x}}{i}{y}) \right),  & \text{if } i \in f_m, \\
    \forall \vect{x} \forall y \left( \vectIndex{x}{i} \leq y \rightarrow f(\vect{x}) \geq f(\subst{\vect{x}}{i}{y}) \right), & \text{if } i \in f_a,
  \end{cases}
\]
and $\vect{x}$ is a vector of $\arity{f}$ variables.

It is easy to see that the $T$-models of $\varphi$ that satisfy the
monotonicity specification $M$ are exactly the $T$-models of
$\varphi^{\mathit{quant}}_M = \varphi \wedge \psi$ (without the
monotonicity specification). In particular, the formula $\varphi$ is
$(T,M)$-satisfiable if and only if $\varphi^\mathit{quant}_M$ is
$T$-satisfiable.

\begin{example}
  Consider again the formula $\varphi$ and the monotonicity
  specification $M$ from Example~\ref{ex:running}. The corresponding formula
  $\varphi^\mathit{quant}_M$ for this monotonicity specification is
  \begin{align}
    \varphi^\mathit{quant}_M = {} &\varphi \wedge \psi^f_1 \wedge \psi^f_2 \wedge \psi^g_1 \\
    = {} &( f(c_1, 2) = 4 ) \wedge (f(c_1+5, 0) = c_2) \wedge (g(c_2) < g(4)) \wedge {} \\
        &\forall x_1 \forall x_2 \forall y . \, (x_1 \leq y \rightarrow f(x_1, x_2) \leq f(y, x_2)) \wedge {} \\
        &\forall x_1 \forall x_2 \forall y . \, (x_2 \leq y \rightarrow f(x_1, x_2) \geq f(x_1, y)) \wedge {} \label{line:antimonotonicity} \\
        &\forall x \forall y . \, (x \leq y \rightarrow g(x) \leq g(y)).
  \end{align}
  This formula precisely encodes the monotonicity specification and is
  $\LIA$-unsatisfiable. The corresponding formula
  $\varphi^\mathit{quant}_{M'}$ for the specification $M'$ does not
  contain the constraint $\psi^f_2$ and its corresponding row
  (\ref{line:antimonotonicity}), which encodes the anti-monotonicity
  of the function $f$ in its second argument. Without this constraint,
  the formula is $\LIA$-satisfiable.
\end{example}

The downside of this approach is that the formula $\varphi^{\mathit{quant}}_M$ is
necessarily quantified for non-trivial monotonicity specifications,
even if the original formula $\varphi$ is quantifier-free. Consequently,
determining its satisfiability might require potentially expensive quantifier
reasoning. Furthermore, while the satisfiability of quantifier-free formulas is
decidable for many theories (such as the theory of uninterpreted functions, \UF),
it is undecidable in general when quantifiers are introduced. For such theories,
the above encoding does not yield a decision procedure for quantifier-free
formulas with monotonicity constraints.

\subsection{Eager Quantifier Instantiation}

Monotonicity can also be expressed without explicit quantification
while preserving satisfiability. In particular, only a finite set of
quantifier instances, so-called \emph{monotonicity lemmas}, are
sufficient to encode the monotonicity specification. Moreover, the set
of these monotonicity lemmas is fixed for each input formula $\varphi$
and can be easily computed from
it~\cite{sofronie2007automated}. However, this approach only works
with the following equivalent definition of monotonicity that
considers all arguments of each function symbol together.

\begin{theorem}
  \label{thm:reformulation}
  For each $\Sigma$-structure $\mathcal{A}$ and monotonicity
  specification $M$, it holds that $\mathcal{A} \models M$ if and only if each
  function symbol $f$ of arity $n$ satisfies $\forall \overline{x}, \overline{y}
  . \, \psi(f, \overline{x}, \overline{y})$ where $\psi(f, \overline{x},
  \overline{y})$ is defined as
  \begin{equation}
    \label{eq:reformulation}
    \left(
      \left ( \bigwedge_{\substack{i \in f_m}} \vectIndex{x}{i} \leq \vectIndex{y}{i} \right ) \wedge
      \left ( \bigwedge_{\substack{i \in f_a}} \vectIndex{y}{i} \leq \vectIndex{x}{i} \right ) \wedge
      \left ( \bigwedge_{\substack{1 \leq i \leq n \\ i \not \in (f_m \cup f_a)}}{\vectIndex{x}{i} = \vectIndex{y}{i}} \right )
    \right )
    \rightarrow
    f(\vect{x}) \leq f(\vect{y}).
  \end{equation}
\end{theorem}

This equivalent definition ensures that the monotonicity lemmas, which
consist of all instances of (\ref{eq:reformulation}) for all pairs of
occurrences of $f$ in $\varphi$, are sufficient to precisely express
the monotonicity specification for any given formula $\varphi$.

\begin{theorem}[\cite{sofronie2007automated,sofronie2014ieee}]
  \label{thm:eager}
  Let $\varphi$ be a formula and $M$ a monotonicity
  specification. Then $\varphi$ is $(T,M)$-satisfiable if and only if
  the formula
  \begin{align*}
    \varphi^{\mathit{inst}}_M ~~\equiv~~ \varphi \wedge \bigwedge_{\substack{f \in \Sigma^F \\ f(\vect{t}), f(\vect{s}) \in \terms{\varphi}}} \psi(f, \vect{t}, \vect{s})
  \end{align*}
  is $T$-satisfiable.
\end{theorem}

Theorem~\ref{thm:eager} yields a decision procedure \solveEager that for a given
formula $\varphi$ and a monotonicity specification computes
$\varphi^\mathit{inst}_M$ and uses an \smt solver for quantifier-free formulas over
the theory $T$. Note that in the worst case, the number of added monotonicity
lemmas $\psi(f, \vect{t}, \vect{s})$ is quadratic with respect to the size of
$\varphi$.

For practical usage of the decision procedure \solveEager, it is
important to note that $\varphi^\mathit{inst}_M$ is not equivalent to
the full encoding of the monotonicity specification
$\varphi^\mathit{quant}_M$; it is merely equisatisfiable with
it. However, the proof in the original
paper~\cite{sofronie2007automated} shows how to algorithmically
construct a $T$-model of $\varphi$ that satisfies the monotonicity
specifications from an arbitrary $T$-model of
$\varphi^\mathit{inst}_M$.

\begin{example}
  Consider again the formula $\varphi$ and the monotonicity specification $M$
  from Example~\ref{ex:running}. Because the set of function applications of $f$
  is $\{ f(c_1, 2), f(c_1 + 5, 0) \}$ and of applications of $g$ is
  $\{ g(c_2), g(4) \}$, the resulting formula $\varphi^\mathit{inst}_M$ is
  \begin{align}
    \varphi^\mathit{inst}_M ~~\equiv~~ {} &( f(c_1, 2) = 4 ) \wedge (f(c_1+5, 0) = c_2) \wedge (g(c_2) < g(4)) \wedge {} \\
      &((( c_1 \leq c_1 + 5) \wedge (2 \geq 0)) \rightarrow f(c_1, 2) \leq f(c_1 + 5, 0)) \wedge {} \label{line:f-app1} \\
      &((( c_1 + 5 \leq c_1) \wedge (0 \geq 2)) \rightarrow f(c_1 + 5, 0) \leq f(c_1, 2)) \wedge {} \label{line:f-app2} \\
      &(( c_2 \leq 4) \rightarrow g(c_2) \leq g(4)) \wedge {} \\
      &(( 4 \leq c_2) \rightarrow g(4) \leq g(c_2)).
  \end{align}
  The formula precisely encodes the monotonicity specification at the points
  where $f$ and $g$ are evaluated and is $\LIA$-unsatisfiable.

  For the monotonicity specification $M'$, which does not enforce
  anti-monotonicity of $f$ in its second argument, the corresponding subformulas
  (\ref{line:f-app1}) and (\ref{line:f-app2}) are changed to
  \begin{align}
    &((( c_1 \leq c_1 + 5) \wedge (2 = 0)) \rightarrow f(c_1, 2) \leq f(c_1 + 5, 0)) \wedge {} \\
    &((( c_1 + 5 \leq c_1) \wedge (0 = 2)) \rightarrow f(c_1 + 5, 0) \leq f(c_1, 2)),
  \end{align}
  which are vacuously true in this case, and the resulting formula
  $\varphi^\mathit{inst}_{M'}$ is $\LIA$-satisfiable. For example, any structure
  $\mathcal{A}$ with $c_1^\mathcal{A} = 6$, $c_2^\mathcal{A} = 0$,
  $f^\mathcal{A}(6, 2) = 4$, $f^\mathcal{A}(11, 0) = 0$, $g^\mathcal{A}(0) = 1$,
  and $g^\mathcal{A}(4) = 2$ is a model of $\varphi^\mathit{inst}_{M'}$. Such a model
  does not need to satisfy all the monotonicity constraints, as the functions
  can behave arbitrarily on points not mentioned by the formula. The proof of
  Theorem~\ref{thm:eager} yields the ``globally monotonized'' version
  $\mathcal{A}^\uparrow$ of $\mathcal{A}$ that interprets the function symbols
  as
  \begin{align*}
    &f^{\mathcal{A}^\uparrow}(x, y) =
      \begin{cases}
        0, &\text{if } x \geq 11 \text{ and } y = 0, \\
        4, &\text{if } x \geq 6 \text{ and } y = 2, \\
        0, &\text{otherwise}, \\
      \end{cases} \\
    &g^{\mathcal{A}^\uparrow}(x) =
      \begin{cases}
        2, &\text{if } x \geq 4, \\
        1, &\text{otherwise}. \\
      \end{cases}
  \end{align*}
\end{example}

\subsection{Lazy Quantifier Instantiation}
\label{sec:lazy-instantiation}

As noted earlier, the number of added monotonicity lemmas
$\psi(f, \vect{t}, \vect{s})$ can be quadratic with respect to the size of
$\varphi$. In many input instances, not all of these monotonicity lemmas are
needed to decide unsatisfiability or to find a model for the formula.
Therefore, we propose a lazy approach that starts by considering only the formula
$\varphi$ and then adds the constraints $\psi(f, \vect{t}, \vect{s})$ lazily,
based on violations of the monotonicity specification. The algorithm is
presented as Algorithm~\ref{alg:lazy}.

\begin{algorithm}
  \caption{Lazy algorithm for solving satisfiability with monotonicity constraints.}
  \label{alg:lazy}
  \begin{algorithmic}[1]
    \Procedure{SolveLazy}{input formula $\varphi$, monotonicity specification $M$}
    \State $\varphi' \gets \varphi$
    \While{$\textsc{IsSAT}(\varphi')$}
    \State $\mathcal{A} \gets \textsc{GetModel}(\varphi')$
    \State $\texttt{violated\_lemmas} \gets \{ \psi(f, \vect{t}, \vect{s}) \mid f \in
    \Sigma^f, f(\vect{t}), f(\vect{s}) \in \terms{\varphi}, \mathcal{A} \not
    \models \psi(f, \vect{t}, \vect{s}) \}$
    \If{$\texttt{violated\_lemmas} = \emptyset$}
    \State \textbf{return} $\textsc{sat}$
    \EndIf
    \State $\varphi' \gets \varphi' \wedge \bigwedge \texttt{violated\_lemmas}$
    \EndWhile
    \State \textbf{return} $\textsc{unsat}$
    \EndProcedure
  \end{algorithmic}
\end{algorithm}

The correctness of Algorithm~\ref{alg:lazy} follows directly from
Theorem~\ref{thm:eager}. In particular, if the algorithm returns \textsc{sat},
the model $\mathcal{A}$ satisfies all instances of $\psi(f, \vect{t}, \vect{s})$,
meaning it satisfies $\varphi^{\mathit{inst}}_M$. Thus, $\varphi$
is satisfiable with the monotonicity specification by
Theorem~\ref{thm:eager}. On the other hand, the \textbf{while} loop maintains the
invariant that $\varphi^{\mathit{inst}}_M \models \varphi'$. Therefore, if the
algorithm returns \textsc{unsat}, the formula $\varphi^{\mathit{inst}}_M$ is not
$T$-satisfiable, which in turn means that $\varphi$ is not $(T, M)$-satisfiable,
again by Theorem~\ref{thm:eager}. Finally, the set of all possible monotonicity
lemmas has size $\mathcal{O}(|\varphi|^2)$. Since at least one violated lemma is
added in each iteration, the algorithm terminates in at most
$\mathcal{O}(|\varphi|^2)$ iterations.

\section{Evaluation}
\label{sec:experiments}

To demonstrate the practical utility of using \smt with monotonicity
constraints for model inference, we perform an extensive evaluation on
discrete models in systems biology, where monotonicity is a
fundamental mechanistic assumption.  Our evaluation focuses on
inferring Boolean~\cite{kauffman1969homeostasis} and
multi-valued~\cite{thomas1991regulatory} networks, two prominent
classes of logic-based
models~\cite{pastva2023repository,kadelka2024meta}.  We have
implemented our procedures within a tool powered by the Z3 \smt
solver~\cite{de2008z3}.  To ensure reproducibility, the tool's source
code is hosted on
GitHub\footnote{\url{https://github.com/sybila/biodivine-algo-smt-inference/tree/artefact-sat-2026-submission}},
while all benchmark instances, test scripts, and complete experimental
results are archived on
Zenodo\footnote{\url{https://doi.org/10.5281/zenodo.18920286}}.

\subsection{Implementation and Tested Tools}
\label{sec:implementation}

We implemented our tool in Rust, utilizing the C bindings for the Z3 \smt solver~\cite{de2008z3}.
The tool implements four distinct \smt encodings:
\begin{itemize}
	\item \textsc{quantified-individual}: Uses the standard encoding from Section~\ref{sec:quantified-encoding}, generating one quantified constraint per monotonic function argument.
	\item \textsc{quantified-aggregated}: Applies Theorem~\ref{thm:reformulation} to generate one aggregated quantified constraint per uninterpreted function.
	\item \textsc{instantiated-eager}: Directly implements the quantifier instantiation from Theorem~\ref{thm:eager}.
	\item \textsc{instantiated-lazy}: Extends the eager approach by lazily introducing monotonicity lemmas, as detailed in Section~\ref{sec:lazy-instantiation}.
\end{itemize}

These encodings are exposed as a generic layer on top of the Z3 \textsc{api}
and can be applied to any uninterpreted function over the \texttt{Int} and \texttt{Bool} sorts.
We then use this framework to implement an inference method for discrete biological models (Section~\ref{sec:inference-encoding}).
For Boolean networks, the encoding relies directly on the \texttt{Bool} sort in Z3.
Meanwhile, for multi-valued networks, the \texttt{Int} sort is used with additional bounds assertions.

For our evaluation, we compare against two state-of-the-art systems biology tools:
Bonesis~\cite{chevalier2024bonesis,chevalier2025data} and AEON~\cite{benevs2022aeon,benevs2023boolean}.
Bonesis utilizes Answer Set Programming (\asp) to explicitly encode unknown functions via a Disjunctive Normal Form (\dnf) representation,
where the maximum number of cubes can be bounded by the user.
AEON, conversely, employs Binary Decision Diagrams ({\bdd}s) to symbolically represent the entire solution space. Note that the \bdd-based approach always computes all solutions and is inherently incapable of computing only a single solution.
We omit Caspo~\cite{guziolowski2013exhaustively}, which has been largely superseded by Bonesis,
and RE:IN~\cite{yordanov2016method,yordanov2023reasoning}, which does not support the full spectrum of monotonic functions.

\subsection{Benchmark Instances}
\label{sec:benchmark-instances}
We evaluate our approach on three classes of benchmark instances.
Two classes, \textsc{bbm-boolean} (8,381 instances) and \textsc{bma-integer} (465 instances),
are derived from established models in the literature.
The third, \textsc{omnipath} (144 instances), represents a real-world,
data-driven inference task utilizing a biological dataset (308 genes, 7 fixed points)
and a dense prior influence graph of known gene regulations (2,570 edges).
A comprehensive description of the benchmark generation process is provided in our Zenodo artifact.
Below, we briefly summarize the methodology for each class.

\paragraph*{Biodivine Boolean Models}
The Biodivine Boolean Models (\bbm) repository~\cite{pastva2023repository} contains 285 Boolean models
curated from the literature and systems biology databases, scaling up to 853 variables and 3,499 regulations.
Most of these models include \emph{constant input variables} representing environmental conditions.
Since realistic inference tasks typically assume a fixed environment,
we generated up to 128 unique random input configurations per model
(exhaustively enumerating all combinations for models with 7 or fewer inputs).
Treating each configuration as a distinct benchmark yielded 8,381 unique problem instances.
We then computed the fixed points for each instance using AEON~\cite{benevs2022aeon},
bounding the enumeration at $2^{16}$ fixed points (a limit hit by only one \bbm model).
The resulting inference task is to recover a model whose influence graph (including monotonicity and essentiality constraints) and fixed-point behavior are equivalent to the original.

\paragraph*{Biomodel Analyzer}
Biomodel Analyzer (\bma)~\cite{bio-model-analyzer} is a framework for multi-valued biological modeling.
We collected 18 publicly available \bma models and applied the same environmental randomization procedure used for \bbm,
yielding 465 problem instances.
Because these instances are multi-valued, their \smt encodings require \texttt{Int} sorts bounded by their respective domain intervals.
The maximum domain interval in this dataset is $\{ 0, \ldots, 10 \}$.

\paragraph*{Omnipath}
Because the \textsc{bbm-boolean} and \textsc{bma-integer} instances are derived from curated models,
they are satisfiable by design and exhibit structurally sparse influence graphs.
To test our approach on a more realistic and challenging scenario,
we constructed an inference task using scRNA-seq observations from seven fully differentiated neural cell types~\cite{DiBella2021}.
This strategy emulates the methodology of recent Boolean network inference case studies~\cite{chevalier2025data,herault2023novel}.
We binarized the raw data using scBoolSeq~\cite{scboolseq} to extract seven partial Boolean fixed-point observations.
We paired these observations with an influence graph of known gene regulations from the Omnipath database~\cite{omnipath},
restricted to a 308-gene strongly connected subnetwork that covers all non-zero genes in the observations.
Crucially, this prior network is dense, featuring a maximum function arity of 76.
However, because we do not enforce regulation essentiality,
the inferred models may be significantly sparser than the prior network.
To simulate varying levels of partial observability,
we randomly sub-sampled the fixed-point observations,
generating 16 variations for each decile between $10\%$ and $90\%$ data retention.
This process produced 144 benchmark instances, roughly $40\%$ of which are satisfiable.

\subsection{Experiment Environment}

We conducted all experiments on a workstation with a 32-core AMD Threadripper 2990WX \textsc{cpu} and 256 GB of \textsc{ram}. All measured times are wall-clock times.
The recorded times include all initialization overhead,
such as parsing the problem specification and generating the \smt query
(or the equivalent data structures for other tools).
The tools were configured to output only the final \textsc{sat}/\textsc{unsat} verdict rather than printing the full inferred model.
We applied a 10-minute timeout to each benchmark instance.
Executions that timed out or crashed (primarily due to out-of-memory errors) were recorded as failures.
Given the high number of benchmark instances, we evaluated up to 16 benchmarks in parallel.

\begin{table*}
	\caption{Summary of successfully solved instances across the three benchmark sets using each \smt encoding (with a 10-minute timeout). 
    }
	\label{tbl:completed-smt-benchmarks}
	\centering
	\setlength\tabcolsep{4 pt}
	\begin{tabular} { L{65pt} R{40pt}  R{60pt}  R{60pt}  R{60pt}  R{60pt}  }
          \toprule
          \multicolumn{1}{c}{} & & \multicolumn{4}{c}{Solved problems} \\
          \cmidrule{3-6}
          \multicolumn{1}{c}{Benchmark set} & Size & Quantified individual & Quantified aggregated & Instantiated eager & Instantiated lazy \\
          \midrule
          {\textsc{bbm-boolean}} & 8381 & 8242 & 7456 & \textbf{8381} & \textbf{8381} \\
          {\textsc{bma-integer}} & 465 & 0 & 0 & \textbf{465} & 417 \\
          {\textsc{omnipath}}    & 144 & 76 & 87 & \textbf{144} & \textbf{144} \\
          \bottomrule
	\end{tabular}
\end{table*}

\subsection{Comparison of SMT Encodings}
\label{sec:comparison-smt}

\begin{figure}[tbp]
	\centering
	\begin{subfigure}[b]{0.49\textwidth}
		\centering
		\includegraphics[width=\textwidth]{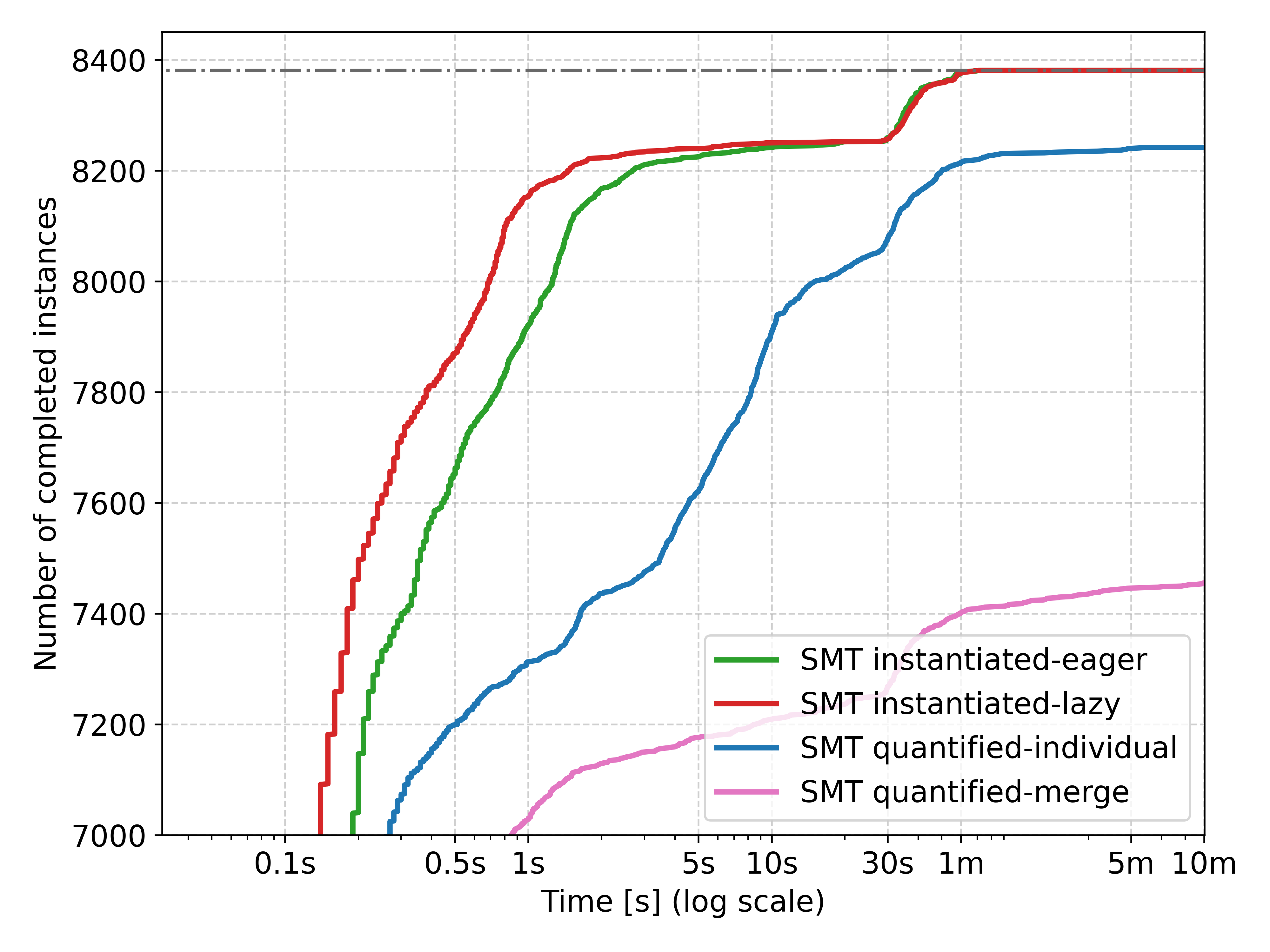}
		\caption{\textsc{bbm-boolean} benchmarks.}
		\label{subfig:smt-bbm-cumulative}
	\end{subfigure}
	\begin{subfigure}[b]{0.49\textwidth}
		\centering
		\vspace{2mm}
		\includegraphics[width=\textwidth]{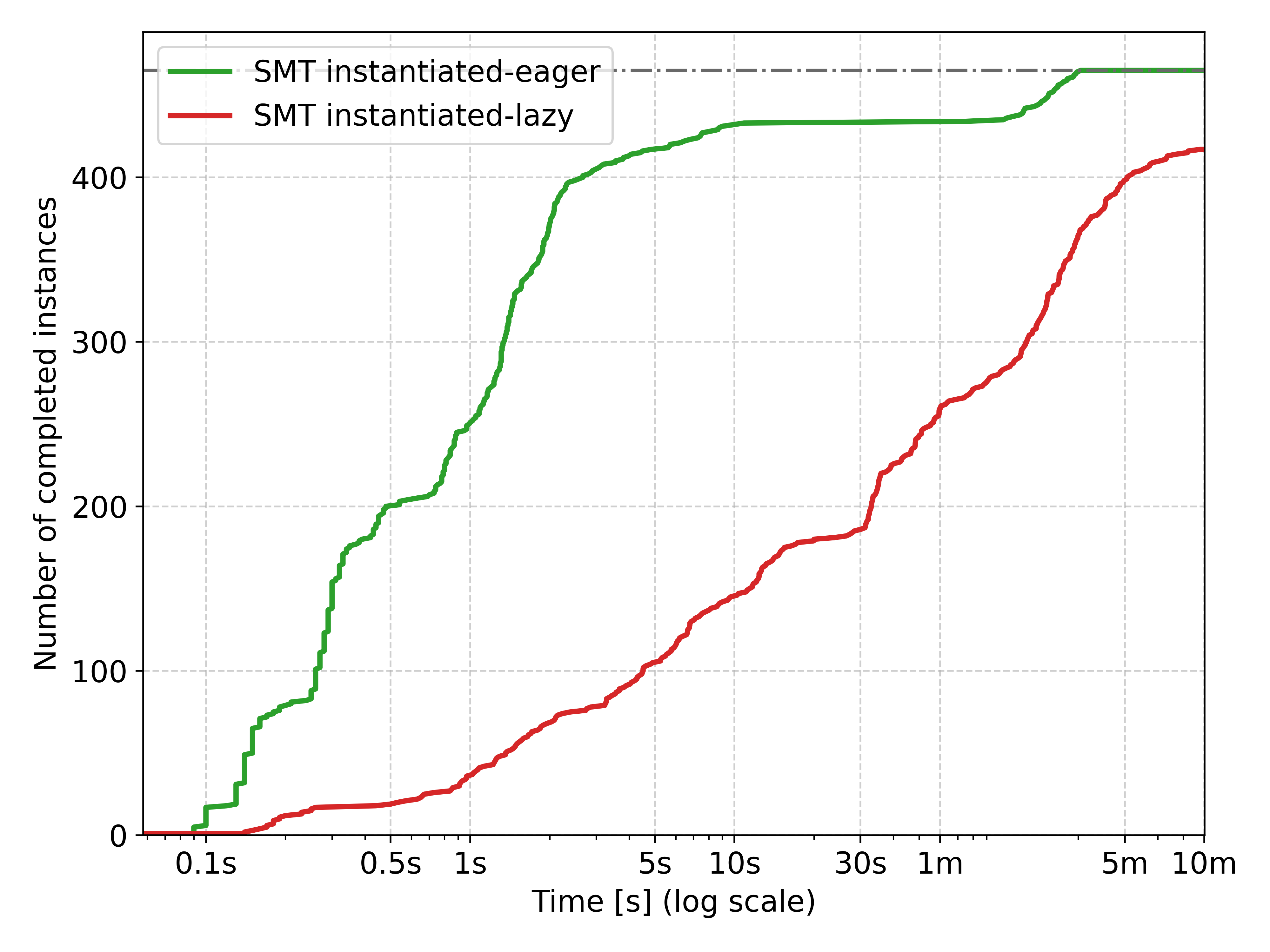}
		\caption{\textsc{bma-integer} benchmarks.}
		\label{subfig:smt-int-cumulative}
	\end{subfigure}
	\begin{subfigure}[b]{\textwidth}
		\centering
		\vspace{2mm}
		\includegraphics[width=0.49\textwidth]{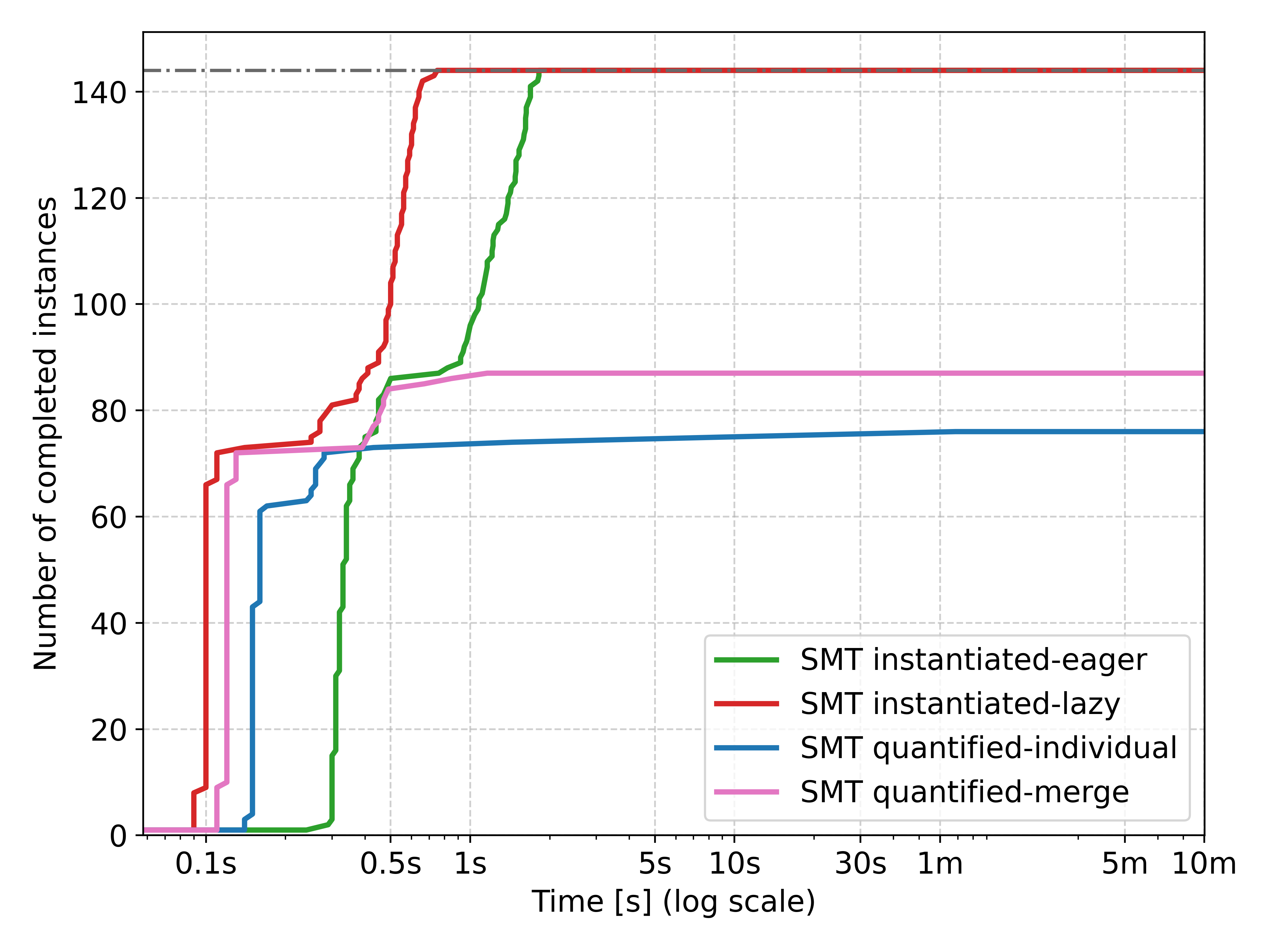}
		\caption{\textsc{omnipath} benchmarks.}
		\label{subfig:smt-omnipath-cumulative}
	\end{subfigure}
	\caption{Cumulative plots comparing the performance of all four implemented \smt approaches on the three benchmark sets. Each plot shows the number of successfully completed benchmark instances by each method (y-axis) before a specific time limit (x-axis). Only methods with at least one successfully solved instance are shown in each plot. The time axis is logarithmic in all plots. For the \textsc{bbm-boolean} plot, the y-axis is truncated to focus on the most challenging instances where the various approaches diverge most significantly.
    }
	\label{fig:smt-cumulative}
\end{figure}

We first evaluated each of the four implemented \smt encodings using the three inference benchmark sets introduced in Section~\ref{sec:benchmark-instances}.
The number of successfully solved instances for each method and benchmark set is summarized in Table~\ref{tbl:completed-smt-benchmarks}.
Both instantiation-based methods successfully completed all \textsc{bbm-boolean} and \textsc{omnipath} instances, with the \textsc{instantiated-eager} method also solving all \textsc{bma-integer} benchmarks. Detailed runtime distributions are shown in the cumulative plots in Figure~\ref{fig:smt-cumulative}, displaying the number of completed instances over time for each monotonicity encoding and benchmark set.

On the \textsc{bbm-boolean} benchmark set, both instantiation-based strategies were able to solve every instance in under 2 minutes.
On the other hand, the quantifier-based strategies failed on 139 and 925 instances, respectively.
To better visualize the performance gaps between the four methods, the plot in Figure~\ref{subfig:smt-bbm-cumulative} is truncated to focus on the most challenging benchmarks where the various approaches diverge most significantly.
High-arity uninterpreted functions proved to be the primary bottleneck for the quantifier-based approaches. While \textsc{quantified-individual} failed on nearly half of the 307 instances where the maximum arity exceeded 12, \textsc{quantified-aggregated} successfully solved only a single instance where the maximum arity exceeded 11.

On the \textsc{bma-integer} benchmarks, the \textsc{instantiated-eager} strategy clearly outperformed the other three approaches.
It was the only method to successfully solve every instance, with all benchmarks completed within 4 minutes and the vast majority finished in under 5 seconds, as illustrated in Figure~\ref{subfig:smt-int-cumulative}.
The \textsc{instantiation-lazy} approach followed closely, solving 417 instances within the 10-minute timeout.
In contrast, the quantifier-based methods did not solve any benchmark within the available time.
This highlights the critical importance of instantiation-based strategies for effectively resolving integer-domain instances.

The performance on the \textsc{omnipath} benchmarks is illustrated in Figure~\ref{subfig:smt-omnipath-cumulative}.
Both the \textsc{instantiated-eager} and \textsc{instantiated-lazy} strategies successfully solved every instance in under 5 seconds.
This demonstrates that the instantiation-based approaches are also well-suited for real-world scenarios involving specifications with high-arity uninterpreted functions.
Notably, \textsc{instantiated-lazy} outperformed \textsc{instantiated-eager} in most cases, highlighting the potential of the lazy approach on sizable, dense networks.
In these scenarios, the eager strategy is forced to generate a large number of monotonicity lemmas that may not actually be necessary to enforce.
By contrast, both the \textsc{quantified-individual} and \textsc{quantified-aggregated} methods failed to solve any satisfiable instance, though they performed well on unsatisfiable instances.
Specifically, the \textsc{quantified-aggregated} variant resolved every unsatisfiable instance in under 3 seconds.
While \textsc{quantified-individual} performed slightly worse, it still successfully resolved all but 11 unsatisfiable instances.

\begin{table*}[tbp]
	\caption{Summary of successfully solved Boolean instances across the approaches compared in Section~\ref{sec:tool-comparison} (with a 10-minute timeout).}
	\label{tbl:completed-tool-benchmarks}
	\centering
	\setlength\tabcolsep{4 pt}
	\begin{tabular} { L{65pt}  R{55pt}  R{55pt}  R{55pt}  R{55pt}  R{55pt}  }
          \toprule
          {Benchmark set} & {Instantiated lazy (\smt)} & {AEON} & {Bonesis (default)} & {Bonesis (16~cubes)} & {Bonesis (8~cubes)}\\
          \midrule
          \textsc{bbm} & \textbf{8381} & 5605 & 7342 & 8246 & 8248 \\
          \textsc{omnipath} & \textbf{144} & 0 & 0 & 0 & 134 \\
          \bottomrule
	\end{tabular}
\end{table*}

\subsection{Comparison Against AEON and Bonesis}
\label{sec:tool-comparison}
Finally, we compared our \smt-based approach to AEON and Bonesis, two state-of-the-art systems biology tools.
Since both of these tools are restricted to Boolean networks, we only consider the \textsc{bbm-boolean} and \textsc{omnipath} benchmark sets.
In particular, we compare our best-performing encoding (\textsc{instantiated-lazy}) against AEON and three Bonesis configurations: the default with no restrictions on the number of allowed \dnf cubes, and two versions where the number of \dnf cubes is restricted to 16 and 8.
Note that by restricting the maximum number of cubes, Bonesis trades completeness for scalability; however, we did not encounter any case where Bonesis returned an incorrect result.
This is in line with the common expectation that biologically motivated update functions
are often simple~\cite{kadelka2024meta}. However, it still means that Bonesis in this configuration can in specific complex instances miss a valid result.

\begin{figure}
    \centering
    \begin{subfigure}[b]{0.49\textwidth}
    	\centering
    	\includegraphics[width=\textwidth]{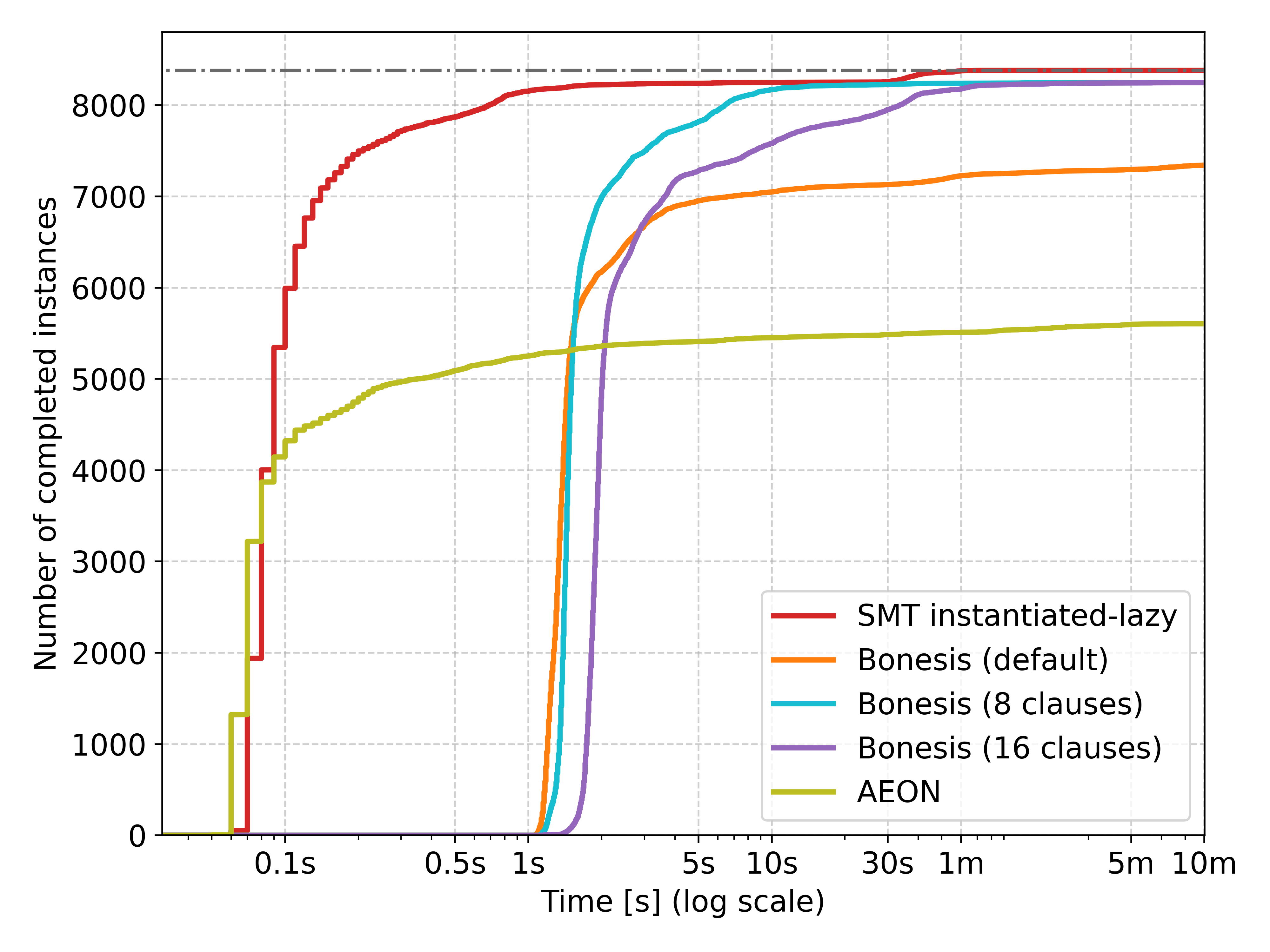}
        \vspace{-2mm}
    	\caption{\textsc{bbm-boolean} benchmarks.}
    	\label{subfig:bbm-all-tools-cumulative}
    \end{subfigure}
    \begin{subfigure}[b]{0.49\textwidth}
    	\centering
        \vspace{2mm}
    	\includegraphics[width=\textwidth]{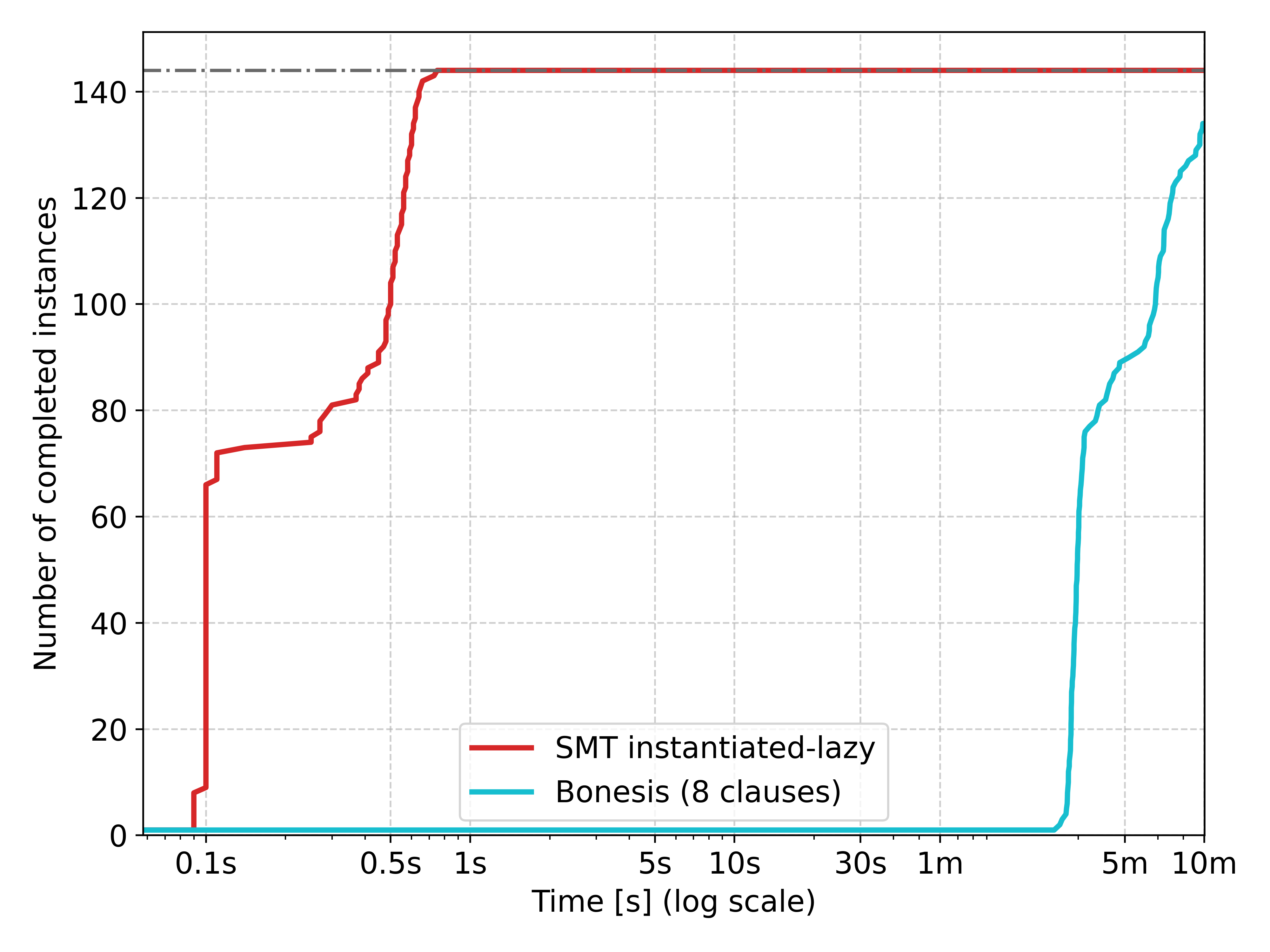}
        \vspace{-2mm}
    	\caption{\textsc{omnipath} benchmarks.}
    	\label{subfig:omnipath-all-tools-cumulative}
    \end{subfigure}
\caption{Cumulative plots comparing the performance of the \textsc{instantiated-lazy} strategy against \texttt{AEON} and three configurations of \texttt{Bonesis} on the two Boolean benchmark sets. Each plot shows the number of successfully completed benchmark instances by each method (y-axis) before a specific time limit (x-axis). Only methods with at least one successful benchmark are shown in each plot. The time axis is logarithmic in all plots.}
\label{fig:all-tools-cumulative}
\end{figure}

The total number of successfully completed instances for each approach and benchmark set is summarized in Table~\ref{tbl:completed-tool-benchmarks}. More details on runtime distributions are shown through the cumulative plots in Figure~\ref{fig:all-tools-cumulative}.

In particular, neither Bonesis nor AEON were able to complete all \textsc{bbm-boolean} instances.
AEON was unable to complete any benchmark where the maximum function arity exceeded 5, failing on close to three thousand instances.
This is an expected result, since AEON always solves the problem exhaustively, computing all satisfiable solutions at once using \bdd encoding.
Bonesis in its default configuration failed to solve over a thousand instances.
On the other hand, the variants restricted to 16 and 8 \dnf cubes performed much better, solving all but 135 and 133 instances, respectively.
Notably, our \textsc{instantiated-lazy} strategy outperformed Bonesis on every single benchmark instance, even when the tool was restricted to only 8 \dnf cubes.
The performance of our \textsc{instantiated-lazy} strategy against the other tools on the \textsc{bbm-boolean} dataset is illustrated through the cumulative plot in Figure~\ref{subfig:bbm-all-tools-cumulative}.

As for the \textsc{omnipath} benchmarks, AEON and the default configuration of Bonesis both failed to handle the high function arity, triggering internal errors during the encoding stage.
Even when restricted to 16 \dnf cubes, Bonesis still failed to successfully solve any instance.
On the other hand, when restricted to 8 \dnf cubes, it was able to complete the majority of the instances.
However, there is an order-of-magnitude difference in computation times between our \textsc{instantiated-lazy} strategy and Bonesis with 8 \dnf cubes.
This difference is illustrated in Figure~\ref{subfig:omnipath-all-tools-cumulative}.

\section{Conclusion}
\label{sec:conclusions}

In this paper, we empirically evaluate the viability of \smt when applied to
the inference of logic-based models in systems biology.
We identify function argument monotonicity as a critical challenge
in the encoding of practical inference problems,
demonstrating that naive quantified encodings of monotonicity frequently fail on real-world instances.
We then evaluate two approaches based on quantifier instantiation,
showing that they significantly outperform the naive quantified encoding.
More importantly, switching to an instantiation-based encoding allows
the \smt approach to outperform current state-of-the-art automated reasoning
tools in systems biology (namely Bonesis and AEON). Furthermore, it enables reasoning over
integer-based systems, a capability that neither of these existing tools supports.

For future work, we aim to extend the inference problem encoding with
additional biologically motivated constraints, such as requirements regarding
trap spaces. Another promising direction is incorporating
the uncertainty of biological observations as soft constraints within the framework
of \smt with optimization.


\bibliography{sat-2026}

\end{document}